**The multilayer connectome of *Caenorhabditis elegans***

Barry Bentley[1], Robyn Branicky[1], Christopher L. Barnes[1], Edward T. Bullmore[2,3] Petra E. Vértes[2†] and William R. Schafer[1†*]

[1)] Neurobiology Division, MRC Laboratory of Molecular Biology, Cambridge UK, [2)] Department of Psychiatry, University of Cambridge, Cambridge UK, [3)] ImmunoPsychiatry, Alternative Discovery & Development, GlaxoSmithKline R&D, Cambridge UK

[†]equal contributions

[*]corresponding author

**Abstract**

Connectomics has focused primarily on the mapping of synaptic links in the brain; yet it is well established that extrasynaptic volume transmission, especially via monoamines and neuropeptides, is also critical to brain function and occurs primarily outside the wired connectome. We have mapped the putative monoamine connections, as well as a subset of neuropeptide connections, in *C. elegans* based on new and published gene expression data. The monoamine and neuropeptide networks exhibit distinct topological properties, with the monoamine network displaying a highly disassortative star-like structure with a richclub of interconnected broadcasting hubs, and the neuropeptide network showing a more recurrent, highly clustered topology. Despite the low degree of overlap between the extrasynaptic and wired networks, we find highly significant multilink motifs of interaction, pinpointing locations in the network where aminergic and neuropeptide signalling modulate synaptic activity. Thus, neuronal connectivity can be mapped as a multiplex network with synaptic, gap junction, and neuromodulatory layers representing alternative modes of interneuronal interaction, providing a prototype for understanding how extrasynaptic signalling can be integrated into a functional connectome.

**Author summary**

Connectomics represents an effort to map brain structure at the level of individual neurons and their synaptic connections. However, neural circuits also depend on other types of interneuronal signalling, such as extrasynaptic modulation by monoamines and peptides. Here we present a draft monoamine connectome, along with a partial neuropeptide connectome, for the nematode *C. elegans*, based on new and published expression data for biosynthetic genes and receptors. We describe the structural properties of these "wireless" networks, including their topological features and modes of interaction with the wired synaptic and gap-junction connectomes. This multilayer connectome of *C. elegans* can serve as a prototype for understanding the multiplex networks comprising larger nervous systems, including the human brain.

## Introduction

The new field of connectomics seeks to understand the brain by comprehensively mapping the anatomical and functional links between all its constituent neurons or larger scale brain regions (Bullmore and Sporns, 2009). The *C. elegans* nervous system has served as a prototype for analytical studies of connectome networks, since the synaptic connections made by each of its 302 neurons have been completely mapped at the level of electron microscopy (Albertson and Thomson, 1976; White et al., 1986). Through this approach, the *C. elegans* nervous system has been found to share a number of topological features in common with most other real-world networks, from human brain networks through social networks to the internet (Bullmore and Sporns, 2009; Stam and Reijneveld, 2007; Varshney et al., 2011). One well-known example is the small-world phenomenon, whereby networks are simultaneously highly clustered (nodes that are connected to each other are also likely to have many nearest neighbours in common) and highly efficient (the average path length between a pair of nodes is short) (Humphries and Gurney, 2008; Watts and Strogatz, 1998). Another characteristic feature of real-world networks which has attracted much attention is the existence of hubs or high-degree nodes, with many more connections to the rest of the network than expected in a random graph (Barabasi and Albert, 1999). As in other networks, these topological features of the *C. elegans* connectome are thought to reflect the functional needs of the system (Bullmore and Sporns, 2012; Vertes and Bullmore, 2015). For example hubs are known to play a privileged role in coordinating functions across a distributed network (Towlson et al., 2013), while the short path lengths (often mediated by the hubs) help increase the efficiency of information transfer across the network (Watts and Strogatz, 1998).

Although connectomics has primarily focused on mapping the synaptic links between neurons, it is well established that chemical synapses are only one of several modes of interaction between neurons. For example, gap junctions, which mediate fast, potentially bidirectional electrical coupling between cells, are widespread in all nervous systems. Likewise, volume transmission and neurohumoral signalling provide means for local or long-range communication between neurons unconnected by synapses. As neuromodulators released through these routes can have profound effects on neural activity and behaviour (Bargmann, 2012; Brezina, 2010; Marder, 2012), a full understanding of neural connectivity requires a detailed mapping of these extrasynaptic pathways.

In *C. elegans*, as in many animals, one important route of neuromodulation is through monoamine signalling. Monoamines are widespread throughout phyla, with evidence that they are one of the oldest signalling systems, evolving at least 1 billion years ago (Walker et al., 1996). In both humans and *C. elegans*, many neurons expressing aminergic receptors are not post-synaptic to releasing neurons, indicating that a significant amount of monoamine signalling occurs outside the wired connectome (Chase and Koelle, 2007). Monoamines are known to be essential for normal brain function, with abnormal signalling being implicated in numerous neurological and psychiatric conditions (Lin et al., 2011). In *C. elegans*, these monoaminergic systems play similarly diverse roles in regulating locomotion, reproduction, feeding states, sensory adaptation, and learning (Chase and Koelle, 2007). Clearly, if the goal of connectomics is to understand behaviourally relevant communication within the brain, extrasynaptic monoamine interactions must also be mapped, not just the network of wired chemical synapses and gap junctions.

In addition to monoamines, neuropeptides are also widely used as neuromodulators in the *C. elegans* nervous system. *C. elegans* contains over 250 known or predicted neuropeptides synthesized from at least 122 precursor genes, and over 100 putative peptide receptors (Hobert, 2013; Li and Kim, 2008). These include homologues of several well-known vertebrate neuropeptide receptors, including those for oxytocin/vasopressin (NTR-1), neuropeptide Y (NPR-1) and cholecystokinin (CKR-2) (Hobert, 2013). As in other animals, neuropeptide signaling is critical for nervous system function, and frequently involves hormonal or other extrasynaptic mechanisms.

This study describes a draft connectome of extrasynaptic monoamine signalling in *C. elegans*, as well as a partial network of neuropeptide signalling, based on new and published gene expression data. We find that the extrasynaptic connectomes exhibit topological properties distinct from one another as well as from the wired connectome. Overall, the neuronal connectome can be modelled as a multiplex network with structurally distinct synaptic, gap junction, and extrasynaptic (neuromodulatory) layers representing inter-neuronal interactions with different dynamics and polarity, and with critical interaction points allowing communication between layers. This network represents a prototype for understanding how neuromodulators interact with wired circuitry in larger nervous systems and for understanding the organisational principles of multiplex networks.

## Results

### A network of extrasynaptic monoamine signalling

To investigate the extent of extrasynaptic signalling in *C. elegans* monoamine systems, we systematically compared the expression patterns of monoamine receptors with the postsynaptic targets of aminergic neurons. Monoamine-producing cells were identified based on the published expression patterns of appropriate biosynthetic enzymes and vesicular transporters (see Methods). The expression patterns for each of the five serotonin receptors (*ser-1, ser-4, ser-5, ser-7* and *mod-1*), three octopamine receptors (*octr-1, ser-3* and *ser-6*), four tyramine receptors (*ser-2, tyra-2, tyra-3* and *lgc-55*), and four of the seven dopamine receptors (*dop-1, dop-2, dop-3* and *dop-4*) were compiled from published data (see Supplemental Tables S1-7).

Three genes previously identified as dopamine-specific receptors were found to have missing or incomplete expression data (*dop-5, dop-6, lgc-55*). To address this, additional expression profiling was conducted for these receptors, using transgenic reporter lines crossed to a series of known reference strains. We observed that all three genes were expressed in small, largely distinct subsets of neurons in the head, body and tail (Figure 1). We were able to identify nearly all the cells with clear reporter expression, giving us a largely completed map of monoamine receptor expression in the *C. elegans* hermaphrodite.

Figure 1. Expression patterns of the dopamine receptors *dop-5, dop-6* & *lgc-53*. Shown are representative images showing expression of GFP reporters under the control of indicated receptor promoters in the head (left panels) or tail/posterior body (right panels). Identified neurons are labelled; procedures for confirmation of cell identities are described in methods. In all panels, dorsal is up and anterior is to the right. In addition to the neurons indicated, dopamine receptor reporters were detected in the following neurons: *dop-5*: BDU (some animals); *lgc-53*: PVPR, CAN (some animals).

Receptor expression patterns suggest that a remarkably high fraction of monoamine signalling must be extrasynaptic. For example, the two tyraminergic neurons, RIML and RIMR, are presynaptic to a total of 20 neurons. Yet of the 114 neurons that express reporters for one or more of the four tyramine (TA) receptors, only 7 are postsynaptic to a tyraminergic neuron (Figure 2a; Supplemental Table S8). Thus, approximately 94% of tyramine-responsive neurons must respond only to extrasynaptic TA. Similar analyses of the other monoamine systems yield comparable results: 100% of neurons expressing octopamine receptors receive no synaptic input from octopamine-releasing neurons (Figure 2b), while 78% of neurons expressing dopamine receptors, and 76% of neurons expressing serotonin receptors receive no synaptic input from neurons expressing the cognate monoamine ligand (Table 1). Thus, most neuronal monoamine signalling in *C. elegans* appears to occur extrasynaptically, outside the wired synaptic connectome. The prevalence of extrasynaptic monoamine signalling between neurons unconnected by synapses or gap junctions implies the existence of a large *wireless* component to the functional *C. elegans* connectome, the properties of which have not previously been studied.

Figure 2. Monoamine signalling in *C. elegans* is primarily extrasynaptic. (A) RIM tyramine releasing neurons, showing outgoing synaptic edges (arrows), and neurons expressing one or more of the four tyramine receptors (grey). (B) RIC octopamine releasing neurons, showing outgoing synaptic edges (arrows), and neurons expressing one or more of the three octopamine receptors (grey). (C) Adjacency matrix showing the synaptic (magenta) and monoamine (green) networks. (D) Multilayer expansion of the synaptic (syn), gap junction (gap), monoamine (MA) and neuropeptide (NP) signalling networks. Node positions are the same in all layers.

**Table 1:** Table showing the number of monoamine receptor-expressing cells that do not receive synapses from releasing cells, and the number of connections in each layer that are non-synaptic, including connections between neurons within the same class

| Network | Non-postsynaptic receptors | | Non-synaptic edges | |
|---|---|---|---|---|
| | № | % | № | % |
| Serotonin | 62 | 75.6 | 457 | 93.3 |
| Dopamine | 137 | 78.3 | 1339 | 96.2 |
| Octopamine | 28 | 100 | 54 | 100 |
| Tyramine | 107 | 93.9 | 216 | 94.7 |
| *Aggregate* | 184 | 74.2 | 2066 | 95.5 |

Using the gene expression data, a directed graph representing a draft aminergic connectome was constructed with edges linking putative monoamine releasing cells (expressing monoamines, biosynthetic enzymes, or transporters) to those cells expressing a paired receptor (Figure 2c; Table 2). Since biologically-relevant long-distance signalling has been experimentally demonstrated in *C. elegans* for both dopamine and serotonin (Chase and Koelle, 2004; Gurel et al., 2012) – while tyramine and octopamine are each released from a single neuronal class (Chase and Koelle, 2007) – edges were not restricted based on the physical distance between nodes. For the serotonin network, only those neurons with strong, consistent expression of serotonin biosynthetic markers such as tryptophan hydroxylase were included (NSM, HSN and ADF).

**Table 2:** Table showing the number of nodes and edges in the individual and aggregate monoamine networks

| Network | Nodes $N$ | № ligand expressing | № receptor expressing | Edges $M^{\rightarrow}$ |
|---|---|---|---|---|
| Serotonin | 86 | 6 | 82 | 490 |
| Dopamine | 175 | 8 | 175 | 1392 |
| Octopamine | 28 | 2 | 28 | 54 |
| Tyramine | 116 | 2 | 114 | 228 |
| *Aggregate* | 248 | 18 | 248 | 2164 |

## The *C. elegans* connectome forms a multiplex network with nonredundant layers

With the inclusion of the monoamine systems, the full *C. elegans* connectome can be considered as a multiplex or multilayer network (Nicosia and Latora, 2015), with each node representing a neuron and each layer of connections – synaptic, gap junction, and monoamine – characterized by distinct edge properties (Figure 2d). For example, chemical synapses represent unidirectional, wired connections that signal on a fast (ms) time scale, while gap junctions generate reciprocal electrical connections that function on an even faster time scale. In contrast, monoamine connections are wireless (with a single sending cell broadcasting to multiple receivers), slow (acting on a time scale of seconds or longer) and unidirectional (Ezcurra et al., 2011; Gurel et al., 2012). Conceptually, additional modes of interneuronal signalling, such as peptide neuromodulation, could represent additional layers.

Prior studies of multiplex networks in non-biological systems – such as communication networks – have tended to find a large degree of overlap between the links observed in distinct layers, implying that they may not be truly independent channels of interaction (De Domenico et al., 2015a).  In contrast, we observe that out of 2036 monoamine connections only 89 overlap with chemical or electrical synapses, meaning 96% of the monoamine connections are unique to the monoamine layer (Figure 2c; Supplemental Table 8). Reducibility analysis (De Domenico et al., 2015a), which clusters the different network layers based on their redundancy or degree of overlap, provides further support that the monoamine networks have a unique structure.  Considered either separately or in the aggregate, the monoamines form a distinct cluster separate from the wired synaptic and gap junction networks (Figure 3a, b). This shows that the monoamine networks overlap less with the synaptic and gap junction networks than the synaptic and gap junction networks do with each other.

Figure 3.  Monoamine networks are largely non-overlapping with the wired connectome. (A-B) Multilayer reducibility dendrograms  Panel A considers monoamine and neuropeptide networks in aggregate; panel B considers monoamine systems individually with neuropeptide systems not included. Layers close on the dendrogram have more overlapping edges and are more reducible.  (C) Degree-degree correlation matrix. Off-diagonal panels show the degree-degree correlation between a pair of network layers. Panels on the diagonal show the degree distribution of the individual layers.  (D) Hive plot showing the wired synaptic and gap junction connections (magenta) and monoamine connections (green). Nodes are classified as sensory, motor or interneurons and are arranged along the three axes according to their degree. Hubs are located further out along the axes.

Similarly, in many previously-described multiplex networks, the high-degree hubs in each layer are often co-located, unequivocally highlighting certain nodes as key controllers of information flow in the system (Nicosia and Latora, 2015). While the synaptic and gap junction layers of the worm connectome are observed to follow this trend, with the same high-degree neurons in both systems (Figure 3c), the extrasynaptic monoamine network exhibits a vastly different structure. While the synaptic and gap junction degrees of individual nodes show high positive correlation (R=.594), no significant degree-degree correlation is observed between the wired and extrasynaptic monoamine layers, indicating that the hubs of the monoamine system are distinct. These analyses suggest two distinct interpretations for the dissimilarity to the wired network layers. Firstly, monoamines may be functioning as an independent network, with little relation to the faster wired network. Secondly, the dissimilarity between layers might indicate that monoamines have a complementary function that is nevertheless coupled to that of the synaptic and gap junction connections.

**Analysis of monoamine network topology**

To address these possibilities, we investigated whether the isolated *C. elegans* monoamine network displays the structural organisation required for information processing. Considered separately, the monoamine networks of *C. elegans* consist of only a few topologically central neurons that broadcast signals to a large number of peripheral neurons. These monoamine-releasing cells are mostly sensory and motor neurons, with the downstream receptors being distributed throughout the worm (Figure 3d). In total, 18 of the 302 neurons in the adult hermaphrodite release monoamines, while 248 neurons (82%) were found to express one or more monoamine receptors. This gives the network a star-like topology, which can be directly observed in all of the separate monoamine networks (Figure 4a). As a consequence, the monoamine network exhibits a heavy tailed distribution containing a small number of high-degree hubs (Figure 3c). This structure is also reflected in other topological network measures, with the monoamine network exhibiting high disassortativity characteristic of star networks (Figure 4b). Disassortativity is known to be relevant in the organisation of collective network dynamics, such as synchronisation (Sorrentino et al., 2006) and cooperation behaviour (Perc et al., 2013; Wang et al., 2008), and is widely observed in other biological and technological networks (Newman, 2006).

Figure 4. Topological properties of the *C. elegans* extrasynaptic networks. (A) Multilayer expansion of monoamine subnetworks. Node positions are the same in all layers.(B-F) Comparison of network metrics for the synaptic (syn), gap junction (gap), monoamine network (MA), aggregate wired & monoamine network (wired+MA), neuropeptide (NP) and complete aggregate (all) networks. Plots show the observed values (filled squares) and expected values for 100 rewired networks preserving degree distribution (boxplots).

The inclusion of these additional monoamine connections into the connectome has a number of effects on the aggregate network. For one, it greatly reduces the overall path length of the network (Figure 4b), increasing the efficiency of integrative information processing by providing paths between more segregated subgraphs of the wired network (Sporns, 2013). In particular, monoamine signalling provides a direct route of communication between sensory neurons and motor neurons (Figure 3d), bypassing the premotor interneurons that play a prominent role in the synaptic and gap junction systems (Towlson et al., 2013). Together, these observations suggest that the monoamines provide efficient global connections for coordinating behaviour throughout the entire organism due to the presence of highly connected hubs directly linking many disparate parts of the network. This is a useful feature given the role of monoamines in signalling physiologically important states relevant to the entire organism, such as food availability (e.g.(Ezcurra et al., 2011)). The increased connectivity provided by the monoamines also results in a reduction in the aggregate network's modular structure, a consequence of increasing the number of connections between functionally segregated units (Figure 4d). The network is, however, still more modular than random, with the monoamine layer also exhibiting greater-than-random modularity compared to null models that rewire the network edges while preserving degree distribution (see Methods). This is expected given the monoamine layer's composition from separate signalling systems.

Due to the hub-and-spoke structure of the extrasynaptic network, the monoamine layer exhibits a much lower level of global clustering (measured here as *transitivity*) than the wired network (Figure 4e). The observation that clustering exists at all is explained by two factors. Firstly, the expression of monoamine receptors by releasing neurons creates a central cluster of hub neurons in the network; secondly, as many neurons also express more than one monoamine receptor, triangles are formed in the network with a receiving neuron as one vertex, and two transmitting neurons as the others. This provides a method of dual lateral inhibition, where a releasing neuron can inhibit antagonistic signals from another hub neuron while simultaneously negating the downstream effects of those signals, a pattern previously observed in the OA/TA and 5-HT systems between RIC/RIM & NSM in the aminergic control of feeding behaviours (Li et al., 2012). Similar patterns also exist within individual monoamine layers; for example, the ventral cord motor neurons express both excitatory (*dop-1*) and inhibitory (*dop-3*) dopamine receptors (Chase et al., 2004), while the expression of an inhibitory receptor (*dop-2*) in dopamine-releasing neurons suggests that the hubs mutually suppress one another to regulate dopamine release.

Many neural and brain networks have been shown to exhibit rich-club organisation (de Reus and van den Heuvel, 2013; Harriger et al., 2012; Shih et al., 2015; van den Heuvel and Sporns, 2011) in which the most highly-connected nodes are more connected to one another than expected by chance (Colizza et al., 2006). It was previously shown that the *C. elegans* wired connectome includes a richclub consisting primarily of a small number of premotor interneurons, controlling forward and backward locomotion (Towlson et al., 2013). Subjecting the monoamine connectome to similar analysis, it was found that this network also contains a distinct richclub (Figure 5a, b; Supplemental Table S10), consisting of dopamine, serotonin, and tyramine-releasing neurons. The rich-club property stems from the fact that most serotonergic neurons contain receptors for both tyramine and dopamine, while dopaminergic and tyraminergic neurons likewise express receptors for the other two aminergic transmitters (Figure 5b), suggesting that the different monoamines coordinate their actions.

Figure 5. Monoamine richclub. (A) Rich-club curve for the directed monoamine network. Dashed line indicates the rich-club coefficient for the *C. elegans* monoamine network and the solid curve is a randomized rich-club curve representing the average rich-club coefficient of 100 random graphs (preserving degree distribution) at each value *k*. Individual rich club neurons are shown in

Table 3 . (B) Schematic showing the separate aminergic systems and the volume transmission signalling between them based on receptor expression. Arrows between boxes denote connections between all of the contained neurons. (C) Connections between the wired & monoamine richclubs. Aminergic rich-club neurons are represented as grey octagons. Members of the wired richclub are shown as circles. Dashed red lines are extrasynaptic links. Solid black lines are chemical or electrical synapses.

**Table 3:** Rich-club neurons of the aggregate monoamine network. Number of neurons in each class are shown in parentheses next to the neuron ID

| Neuron ID | Degree $k_{ma}$ | Rich-club $\Phi_{norm}$ | MA | Receptors | Type |
|---|---|---|---|---|---|
| CEP (4) | 185 | 3σ | DA | *dop-2*, *octr-1*, *tyra-3* | Sensory |
| ADE (2) | 185 | 3σ | DA | *dop-2*, *octr-1*, *tyra-3* | Sensory |
| PDE (2) | 181 | 3σ | DA | *dop-2* | Sensory |
| RIM (2) | 128 | 3σ | TA | *ser-4*, *mod-1*, *dop-1* | Motor |
| NSM (2) | 96 | 3σ | 5-HT | *ser-4*, *dop-3*, *ser-2*, *tyra-2* | Pharynx |
| HSN (2) | 92 | 1σ | 5-HT | *lgc-53*, *lgc-55* | Motor |

**Properties of a partial neuropeptide network**

We next investigated the structure of the signalling network for neuropeptides. Since the receptors for many neuropeptides, and the ligands for many neuropeptide receptors, remain unknown, we therefore generated a partial neuropeptide network whose properties we could compare to the synaptic, gap junction and monoamine networks. We focused on 12 neuropeptide receptors with well-established ligands (with biologically-plausible EC50 values in in vitro assays) and precisely-characterized expression patterns for both receptor and peptide precursor genes (Supplemental Tables S11-12). Networks were classified by receptor, allowing many-to-many relationships between neuropeptides and receptors. Even for this partial network, 239 neurons are seen to be involved in neuropeptide signalling (out of 302 possible) with 7046 connections between them, providing greater connectivity than either the synaptic or monoamine layers. Of the receptor-expressing neurons, almost 60% received no synaptic input from neurons expressing one of their ligands, suggesting that neuropeptide signalling, like monoamine signalling, is largely extrasynaptic. Likewise, the majority of edges in the neuropeptide network do not overlap with synapses (97% non-overlapping), again consistent with a largely extrasynaptic mode of signalling for neuropeptides (Figure 6a).

Figure 6. Neuropeptide networks. (A) Adjacency matrix showing the synaptic (magenta) and neuropeptide (green) networks. (B) Multilayer reducibility dendrograms for individual neuropeptide networks. Layers close on the dendrogram have more overlapping edges and are more reducible. Wired and monoamine layers are italicized and indicated with green (MA), blue (gap junction), or magenta (synaptic) boxes. (E) Multilayer expansion of wired, monoamine, and neuropeptide networks. Node positions are the same in all layers.

The neuropeptide network, like the monoamine network, exhibits a structure distinct from the wired connectome. No significant degree correlation was observed between the partial neuropeptide network and the synaptic, gap junction, or monoamine networks, indicating that neuropeptide hubs are distinct from those in other layers (Figure 3c). Likewise, reducibility analysis shows low overlap between the neuropeptide edges and those in the monoamine, synaptic and gap junction layers (Figure 3a). Interestingly, some individual neuropeptide systems, in particular CKR-2, overlap significantly with the networks of monoamine systems, while others, including the neuropeptide F/Y receptors NPR-1/2/5/11, show little overlap with either the wired or other extrasynaptic networks (Figure 6b).

Examining the network measures for the neuropeptide network reveal it to have some topological properties in common with the monoamine network, but also crucial differences. For example, both networks have a shorter characteristic path length and lower modularity than the wired networks (Figure 4c, d). On the other hand, the neuropeptide network has much higher clustering than any other connectome layer (Figure 4e), and is significantly less disasssortive (Figure 4b) than the monoamine network. In part, this is an expected consequence of the large number of connections in the neuropeptide network; however, the observed clustering in the neuropeptide network was significantly higher even than null models with the same edge density. In addition, the neuropeptide network shows much higher reciprocity than the monoamine network (Figure 4f), with the individual neuropeptide systems generally lacking the star-like topology characteristic of the monoamines (Figure 6c).

**Modes of interaction between wired and extrasynaptic layers**

Despite the distinct structures and topologies of the different neuronal connectome layers, they are likely to interact in functionally significant ways. For example, although the wired and monoamine richclubs do not overlap, there are significant links between them (Figure 5d). To systematically identify neurons that have a role in linking all of the layers, neurons were first ordered according to the product of their degree-rank across the synaptic, gap junction and monoamine layers (Table 1a). We observe that the five highest ranking neurons, which have the highest participation across all layers, include two from the monoamine richclub (RIML and RIMR) and two from the wired rich club (RIBL, and DVA). Indeed, the premotor interneuron DVA is a receiver for serotonin, dopamine and tyramine signalling, while the tyraminergic RIMs are highly connected to the premotor interneurons of the wired rich club (Figure 5c, 7a-b). As one might expect from their topological role in linking the monoamine and wired network layers, the RIMs have been shown in a number of studies to play a central role in the modulation of sensory pathways in response to feeding states as well as the control of downstream locomotion motor programs (Donnelly et al., 2013; Piggott et al., 2011; Wragg et al., 2007). Similarly RIB, which expresses receptors for serotonin and dopamine, is thought to integrate numerous sensory signals (Mori and Ohshima, 1995; Tsalik et al., 2003) and has been demonstrated to influence reorientation in foraging behaviour (Gray et al., 2005).

Figure 7. Modes of interaction between connectome layers: (A-B) Multilayer diagram showing the connections made by the DVA (A) and RIM (B) neurons. Node positions are the same in all layers. (B-C) Motif *z*-scores for monoamines (B) and neuropeptides (C). Over-represented motifs are represented by red upward-pointing triangles. Under-represented motifs are represented by blue downward-pointing triangles. Non-significant motifs are shown by black squares. Values for randomized null model networks are shown as grey crosses. Asterisks report the significance level: * indicates $p \leq 0.05$ ; ** indicates $p \leq 0.01$; **** indicates $p \leq 0.0001$. Examples of monoamine motif 10 and neuropeptide motif 20 are listed in Tables 5 and 6. (E) Multilink motif IDs. These correspond to all possible configurations of links between two neurons allowing for: no connection of a given type (dotted line), directed extrasynaptic monoamine links (Ext, represented as arrows on the top), bidirectional gap junctions (represented as bars in the middle) and synapses (represented as inverted arrowheads on the bottom line).

**Table 4 Multilayer hub neurons for 3-layer and 4-layer connectomes.** The normalized degree product ($k_{norm}$) showing the neurons with the highest degree rank across all of the layers. Rich-club neurons are indicated with ★

**Table 4a: 3-layer (syn, gap, MA) normalized degree product**

| Neuron | $k_{norm}$ | $k_{syn}$ | $k_{gap}$ | $k_{ma}$ |
|---|---|---|---|---|
| RIMR★ | 0.238 | 34 | 14 | 128 |
| RIBL★ | 0.194 | 29 | 30 | 14 |
| RIML★ | 0.173 | 28 | 12 | 128 |
| RIBR | 0.167 | 25 | 30 | 14 |
| DVA★ | 0.129 | 54 | 10 | 16 |
| RIS | 0.111 | 27 | 16 | 14 |
| PVT | 0.095 | 27 | 12 | 16 |
| ADEL★ | 0.073 | 31 | 4 | 185 |
| RICL | 0.068 | 23 | 8 | 44 |
| VD01 | 0.066 | 14 | 16 | 16 |

**Table 4b: 4-layer (syn, gap, MA, NP) normalized degree product**

| Neuron | $k_{norm}$ | $k_{syn}$ | $k_{gap}$ | $k_{ma}$ | $k_{np}$ |
|---|---|---|---|---|---|
| RIMR★ | 0.166 | 34 | 14 | 128 | 114 |
| RIML★ | 0.121 | 28 | 12 | 128 | 114 |
| DVA★ | 0.085 | 54 | 10 | 16 | 104 |
| PVQR | 0.044 | 22 | 10 | 16 | 110 |
| ASHR | 0.043 | 21 | 12 | 10 | 162 |
| RIS | 0.034 | 27 | 16 | 14 | 44 |
| PVT | 0.032 | 27 | 12 | 16 | 48 |
| ADFR | 0.031 | 21 | 4 | 90 | 162 |
| VD01 | 0.031 | 14 | 16 | 16 | 61 |

Multilink motif analysis provides another approach for investigating the interactions between the synaptic, gap junction and monoamine layers (Menichetti et al., 2014). Since each layer contains the same set of nodes but a different pattern of edges, the frequencies with which different combinations of links co-occur between pairs of nodes throughout the multiplex network can be determined. Of the 20 possible multilink motifs, seven were found to be overrepresented and four underrepresented compared to networks composed from randomized layers (Figures 7c). Several of these do not involve monoamines; for example, *motif 1*, containing no connections, indicates that specific link combinations are more likely to co-occur than expected by chance, therefore also increasing the number of 'empty' pairs compared to the randomized case. Three additional overrepresented motifs – reciprocal chemical synapses (*motif 3*) and the co-occurrence of a gap junction with a single or reciprocal chemical synapse (*motifs 5 & 6*) – have been reported in an earlier analysis of the wired network (Varshney et al., 2011). These also align with results from the degree-degree correlation and reducibility (Figures 3a, b) indicating that synapses and gap junctions frequently overlap. This is mirrored in the underrepresentation of *motifs 2 & 4* corresponding to synapses or gap junctions alone.

Although the overlap between monoamine and wired connectivity is low, multilink motif analysis revealed a few overrepresented motifs involving monoamines. The most interesting (and statistically significant) of these corresponds to a unidirectional monoamine link coincident with reciprocal synaptic connections (*motif 10*). The structure of this motif is well-suited to provide positive or negative feedback in response to experience, suggesting that this may be a functionally important aspect of monoamine activity within the wider network. Indeed, connections of this type (Table 4a) have been implicated in a number of *C. elegans* behaviours; for example, *motif 10* connections between ADF and AIY have been shown to be important for the learning of pathogen avoidance (Zhang et al., 2005) and connections between RIM and RMD are important for the suppression of head movements during escape behaviour (Pirri et al., 2009). *Motif 10* connections between PDE and DVA are also thought to play a role in controlling neuropeptide release (Bhattacharya et al., 2014).

**Table 5:** Examples of monoamine multilink *motif 10*. List of neurons connected by *motif 10* (i.e. unidirectional MA link, no gap junctions, and reciprocal synapses)

| Cell A | | Cell B |
|---|---|---|
| NSM (L/R) | → | I6 |
| ADFR | → | ASHR |
| ADFR | → | AWBR |
| CEP (DL/VL) | → | OLLL |
| CEP (DR/VR) | → | OLLR |
| ADEL | → | IL2L |
| ADE (L/R) | → | FLPL |
| ADER | → | FLPR |
| ADFR | → | AIYR |
| ADEL | → | BDUL |
| CEPDR | → | RIS |
| PDEL | → | DVA |
| RIMR | → | RMDR |
| HSNL | → | AIAL |
| HSNR | → | AVJL |
| HSNR | → | PVQR |

Interestingly, although the neuropeptide network showed little structural overlap with the monoamine network, its modes of interaction with the wired connectome showed striking parallels. When the neuropeptide network was included in the multiplex participation analysis, we observed that the RIM and DVA neurons continue to play central roles in linking the four network layers (Figure 6a, b; Table 4b). Likewise, multilink motif analysis, this time using the neuropeptide and wired layers, again identified *motif 10* (a unidirectional neuromodulatory connection coincident with a reciprocal synaptic connection) as significantly overrepresented, further supporting the notion that this motif plays a key role in extrasynaptic modulation of synaptic computation (Figure 7d). Even more highly overrepresented relative to expectation was *motif 20*, reciprocal neuropeptide and synaptic connections coincident with a gap junction. This motif was not overrepresented in the multilinks analysis for monoamines, perhaps because of the low reciprocity of the monoamine network. Interestingly, several of the *motif 20* multilinks (Supplemental Table S15) are components of the RMG hub and spoke network, which has been implicated in the control of various behaviours including locomotion, aggregation, and pheromone response (Jang et al., 2012; Macosko et al., 2009).

**Table 6:** Examples of neuropeptide multilink *motif 20*. List of neurons connected by *motif 20* (i.e. reciprocal NP link, gap junction, and reciprocal synapses)

| Cell A | | Cell B |
|---|---|---|
| FLPL | ↔ | FLPR |
| PHAL | ↔ | PHAR |
| PHBL | ↔ | PHBR |
| RMGL | ↔ | URXL |
| RMGR | ↔ | URXR |
| RMGR | ↔ | ASHR |
| PVR | ↔ | DVA |
| AVAL | ↔ | AVAR |

**Discussion**

This study has analysed the properties of an expanded *C. elegans* neuronal connectome, which incorporates newly-compiled networks of extrasynaptic monoamine and neuropeptide signalling. Analyses reveal that these extrasynaptic networks have structures distinct from the synaptic network, and from one another. The monoamine network has a highly disassociative, star-like topology, with a small number of high-degree broadcasting hubs interconnected to form a rich-club core. The monoamine systems are thus well-suited to broadly coordinate global neural and behavioural states across the connectome.

In contrast, the neuropeptide network shows a highly clustered topology with substantially higher reciprocity, suggesting the importance of neuropeptide systems in the cohesion and functional segregation of nervous system modules. While these extrasynaptic networks are separate and non-overlapping with the wired connectome, the hubs of both the wired and wireless networks are interconnected, with multilink motifs showing interaction between the systems at specific points in the network. This suggests that the extrasynaptic networks function both independently – coordinating for example through the monoamine richclub – and in unison with the synaptic network through multilayer hubs such as RIM and DVA and through overrepresented multilink motifs.

The importance of extrasynaptic neuromodulation to the function of neural circuits is clearly established, for example from work on crustacean stomatogastric circuits (Marder, 2012). However, systematic attempts to map whole-organism connectomes have focused primarily on chemical synapses, with even gap junctions being difficult to identify using high-throughput electron microscopy approaches (Chklovskii et al., 2010). The incorporation of extrasynaptic neuromodulatory interactions, inferred here from gene expression data, adds a large number of new links largely non-overlapping with those of the wired connectome. Although the valence and strength of these inferred neuromodulatory links are largely unknown (information also lacking for much of the synaptic connectome), the monoamine and neuropeptide networks described here nonetheless provide a far more complete picture of potential pathways of communication between different parts of the *C. elegans* nervous system.

**Topological properties of monoamine and neuropeptide networks**

Although monoamine and neuropeptide signalling both occur extrasynaptically and act on similar timescales, the monoamine and neuropeptide networks show distinct topologies, perhaps reflecting differences in biological function. As noted previously, the monoamine network has a star-like architecture that is qualitatively different to the other network layers. This structure is reflected in the network's high disassortativity and in the low number of recurrent connections. In addition, we observed that the monoamine network contains a rich club of highly interconnected high-degree releasing neurons, whose members are distinct from (though linked to) the rich club of the wired connectome. Together, this structure is well-suited to the organisation of collective network dynamics, and is a useful feature given the role of monoamines in signalling physiologically important states relevant to the entire organism, such as food availability.

Despite enormous differences in scale, the monoamine systems of *C. elegans* and mammals share a number of common properties suggestive of common network topology. As in *C. elegans*, mammalian brains contain a relatively small number of monoamine-releasing neurons that project widely to diverse brain regions; for example, in humans serotonin is produced by less than 100,000 cells in the raphe nuclei, or one millionth of all brain neurons (Trueta and De-Miguel, 2012). Moreover, extrasynaptic volume transmission is thought to account for much, if not most, monoamine signalling throughout the mammalian brain (De-Miguel and Fuxe, 2012; Fuxe et al., 2012). Parallels between monoamine systems in *C.*

*elegans* and larger nervous systems are not exact; for example, in *C. elegans*, most if not all aminergic neurons appear capable of long-distance signalling, whereas monoamines in larger nervous systems can be restricted by glial diffusion barriers (Owald and Waddell, 2015). Nonetheless, mammalian monoamine-releasing neurons, like their *C. elegans* counterparts, appear to function as high-degree broadcasting hubs with functionally and spatially diverse targets (Trueta and De-Miguel, 2012). Thus, understanding how such hubs act within the context of the completely mapped wired circuitry of *C. elegans*, may provide useful insights into the currently unknown structures of multilayer neuronal networks in larger animals.

Although the neuropeptide network has been only partially characterized, its structure clearly differs in important ways from the other connectome layers, including the monoamine network. In particular, the neuropeptide layer shows strikingly high clustering, even taking into account its high density of connections, and much higher reciprocity than the monoamine network. These properties suggest the neuropeptide networks are important for segregation and cohesiveness of functional modules within the nervous system.

Multilinks analysis also identified differences between the extrasynaptic monoamine and neuropeptide networks. In both cases, a unidirectional extrasynaptic connection coincident with a recurrent synaptic connection (motif 10) was overrepresented in the multiplex connectome. This motif is well-suited to provide feedback between linked nodes, and occurs in several microcircuits implicated in learning and memory. For neuropeptides, a second multilink motif, involving reciprocal neuromodulatory and synaptic connections coincident with a gap junction (motif 20) was even more highly overrepresented. This motif occurs in several places in the RMG-centred hub-and-spoke circuit that plays a key role in control of aggregation and arousal. As more neuropeptide systems become characterized, it is reasonable to expect additional examples of this motif will be identified; these may likewise have important computational roles in key neural circuits.

**A prototype for multiplex network analysis**

While network theory has occasionally provided novel insights in *C. elegans* biology, more often the *C. elegans* wired connectome has provided a useful test-bed for validating new network theoretical concepts or their application to larger mammalian brains (Vertes and Bullmore, 2015). In recent years, multilayer complex systems have become an area of intense focus within network science, with a large number of papers dedicated to extending classical network metrics to the multilayer case and to developing new frameworks to understand the dynamical properties of multilayer systems (Kivelä et al., 2014).

By definition, multilayer networks contain much more information than simple monoplex networks, leading to significant data-collection challenges. In social networks, for example, large monoplex datasets have been collected describing various types of interactions between people, but these are typically disparate datasets based on different populations. Multiplex datasets combining various edge types into a number of layers are often restricted in size (the number of nodes for which data are collected) or in the choice of edges it is possible to consider (interaction types constrained by data availability) (Kivelä et al., 2014).

The multiplex connectome of *C. elegans* has the potential to emerge as a gold standard in the study of multilayer networks, much like the wired *C. elegans* connectome has for the study of simple monoplex networks over the last 15 years. The synaptic, gap junction, and monoamine layers already represent a relatively reliable and complete mapping of three distinct connection types. The lack of degree-degree correlation between some of these layers suggests that they are not just different facets of one true underlying network but rather distinct channels of communication, which are likely coupled in higher order motif structures. The different time-scales on which each of the layers operates are also likely to allow the emergence of interesting dynamical phenomena. Finally, the large number of distinct extrasynaptic interactions offers the scope for a more refined dataset, each aligned to the same complete set of 302 nodes.

**Prospects for complete mapping of multilayer connectomes**

How feasible is it to obtain a complete multiplex neuronal connectome? Although the neuropeptide network described here represents only a sample of the total network, the monoamine network already represents a reasonable draft of a complete monoamine connectome. Since expression patterns for amine receptors have been based on reporter co-expression with well-characterized markers, the rate of false positives (i.e. neurons falsely identified as monoamine receptor expressing) is probably very low. In contrast, the false-negative rate (monoamine receptor-expressing cells not included in the network) is almost certainly somewhat higher. In some cases (e.g. *dop-4* and *dop-3* in ASH (Ezak and Ferkey, 2010; Ezcurra et al., 2011)), reporter transgenes appear to underreport full functional expression domains; in others (e.g. *ser-5*) only a subset of cells expressing a particular reporter have been identified (Harris et al., 2009). With recently developed marker strains (Pereira et al., 2015; Serrano-Saiz et al., 2013), it should be possible to revisit cell identification and fill in at least some of these missing gaps. In addition, other monoamines (e.g. melatonin (Tanaka et al., 2007)) might function as neuromodulators in *C. elegans*, and some of the currently uncharacterized orphan receptors in the worm genome (Hobert, 2013) might respond to monoamines. Potentially, some of these receptors might be expressed in postsynaptic targets of aminergic neurons (in particular, those of dopaminergic and octopaminergic neurons, which are not known to express classical neurotransmitters). However, the existence of additional monoamine receptor-expressing cells also means that non-synaptic edges are almost certainly undercounted in the network. Thus, the high degree of monoamine releasing hubs – and their importance for interneuronal signalling outside the wired connectome – is if anything understated by the current findings.

In the future, it should be possible to expand the scope of the multilayer connectome to gain a more complete picture of interneuronal functional connectivity. Obtaining extrasynaptic connectomes for larger brains, especially those of mammals, will likely be vastly more complicated than for *C. elegans*, due not only to the increase in size, but also the existence of additional structural and dynamical properties, such as glial barriers, cellular swelling, and arterial pulsations, all of which dynamically alter extracellular diffusion (Sykova and Nicholson, 2008; Taber and Hurley, 2014). In contrast, reanalysis of reporters for monoamine receptors using recently developed reference strains (Pereira et al., 2015; Serrano-Saiz et al., 2013) could provide a largely complete monoamine signalling network for *C. elegans*. A greater challenge would be to obtain a complete neuropeptide network; this would require comprehensive de-orphanization of neuropeptide GPCRs as well as expression patterns for hundreds of receptor and peptide genes. Additional layers of neuronal connectivity also remain unmapped, such as extrasynaptic signalling by insulin-like peptides, purines, and classical neurotransmitters such as acetylcholine and GABA (Chan et al., 2013; Dittman and Kaplan, 2008; Pierce et al., 2001). Obtaining this information, while difficult, is uniquely feasible in *C. elegans* given the small size and precise cellular characterisation of its nervous system. Such a comprehensive multilayer connectome could serve as a prototype for understanding how different modes of signalling interact in the context of neuronal circuitry.

**Materials & Methods**

*Synaptic & gap junction networks*

The synaptic and gap junction networks used in this work were based on the full hermaphrodite *C. elegans* connectome, containing all 302 neurons. This network was composed from the somatic connectome of White et al (White et al., 1986), updated and released by the Chklovskii lab (Chen et al., 2006; Varshney et al., 2011); and the pharyngeal network of Albertson and Thomson (Albertson and Thomson, 1976), made available by the Cybernetic *Caenorhabditis elegans* Program (CCeP) (http://ims.dse.ibaraki.ac.jp/ccep/) (Oshio et al., 2003). The functional classifications referred to in the text (i.e. *sensory neuron, interneuron, motoneuron*) are based on the classification scheme used in WormAtlas (Altun et al., 2002).

*Monoamine network construction*

To map the aminergic signalling networks of *C. elegans*, a literature search was first performed to identify genes known to be receptors, transporters or synthetic enzymes of monoamines. A further search was performed to collect cell-level expression data for the monoamine associated genes identified in the previous step. This search was assisted with the curated expression databases of WormBase (Version: WS248; http://www.wormbase.org/) (Howe et al., 2016) and WormWeb (Version date: 2014-11-16)(Bhatla, 2014). A summary of these data is in Supplemental Tables S1-S7. Neurons expressing multiple receptors for a single monoamine receive a single edge from each sending neuron. Reciprocal connections between nodes are considered as two separate unidirectional connections. Edge lists for individual network layers along with a data model showing the relations and constraints used to construct the network are provided in the SI. For completeness, a version of the serotonin network including the weakly expressing neurons is also included.

*Neuropeptide network construction*

The neuropeptide network was constructed from published expression data for peptides and receptors, using an approach similar to that used for the monoamines. Only those systems were included for which sufficient expression and ligand-receptor interaction data existed in the literature, with interactions being limited to those with biologically plausible peptide-receptor $EC_{50}$ values (shown in Figure 7b). In total, 15 neuropeptides and 12 receptors were matched and included in the network. Networks were classified by receptor, allowing a many-to-many relationship between neuropeptides and receptors.

*Neuron identification & microscopy*

The expression patterns of the dopamine receptors were determined using the reporter strains DA1646 *lin-15B & lin-15A(n765) X; adEx1646 [lin-15(+) T02E9.3(dop-5)::GFP]*, BC13771 *dpy-5(e907) I; sEX13771 [rCesC24A8.1(dop-6)::GFP + pCeh361]*, and FQ78 *wzIs26 [lgc-53::gfp; lin-15(+)];lin-15B & lin-15A(n765)* (kindly provided by Niels Ringstad).

The neurons expressing the receptors were identified based on the position and shape of the cell bodies and in most cases co-labelling with other markers. The reporter strains were all crossed with the cholinergic reporter (Pereira et al., 2015) OH13646 *pha-1(e2123) III; him-5(e1490) V; otIs544 [cho-1(fosmid)::SL2::mCherry::H2B + pha-1(+)]* and the glutamatergic reporter (Serrano-Saiz et al., 2013)OH13645 *pha-1(e2123) III; him-5(e1490) V; otIs518 [eat-4(fosmid)::SL2::mCherry::H2B + pha-1(+)]* (both kindly provided by Oliver Hobert), and dye-filled with DiI using standard procedures . Strains were also crossed to AQ3072 *ljEx540[cat-1::mcherry]* and PT2351 *him-5(e1490) V; myEx741 [pdfr-1(3kb)::NLS::RFP + unc-122::GFP]*, which label cells expressing the vesicular monoamine transporter and the PDFR-1 receptor, respectively. When ambiguous, reporter strains were crossed with additional strains, as listed below.

Reporter expression in individual neurons was confirmed with the following crosses:

For *dop-5*:

AIM and ADF were confirmed based on coexpression with *cat-1*. RIB, AIY, M5, and DVA were identified based on position and coxpression with *cho-1*(Pereira et al., 2015). MI, ASE (previously identified in (Etchberger et al., 2009)) and ADA were confirmed based on position and coexpression with *eat-4* (Serrano-Saiz et al., 2013). PHA and PHB were confirmed based on costaining with DiI. PVT and BDU were identified based on cell body position and shape alone.

For *dop-6*:

RIH and ADF were confirmed based on coexpression with *cat-1*(Duerr et al., 1999). ASI and PHA were confirmed based on costaining with DiI. AQ3499 *ljEx805 [sra-6::mcherry + PRF4]* was used to confirm expression in PVQ. AQ3682 *ljEx921[flp-8::mcherry + unc-122::gfp]* was used to confirm expression in URX. IL2, RIB, and URA were identified based on position and coexpression with *cho-1*. OLL (previously identified in (Smith et al., 2010)) was identified based on position and coexpression with *eat-4*. AVF was identified based on position and failure to coexpress *eat-4* and *cho-1*.

For *lgc-53*:

AIM was confirmed based on coexpression of *cat-1*. AVF was confirmed based on coexpression with *pdfr-1* and failure to coexpress *eat-4* and *cho-1*. URY was confirmed based on position, coexpression with *eat-4*, and lack of coexpression with *ocr-4*. AQ3526 *ljEx822 [klp-6::mcherry + pRF4]* was used to confirm IL2 expression. AQ3535 *ljEx828 [unc-4::mcherry + pRF4]* was used to confirm VA expression. PVPR was confirmed based on position and coexpression with *cho-1*.

FLP was confirmed based on position, morphology, and coexpression with *eat-4*. HSN, CAN and PVD expression were identified based on position and morphology.

*Microscopy*

Strains were examined using a Zeiss Axioskop. Images were taken using a Zeiss LSM780 confocal microscope. Worms were immobilized on 3% agarose pads with 2.5mM levamisole. Image stacks were acquired with the Zen 2010 software and processed with Image J.

*Topological network measures*

Edge counts, adjacency matrices and reducibility clusters were all computed using binary directed versions of the networks. The same networks, excluding self-connections (**Tr**(A) = 0), were used to compute all other measures.

Network measures are compared to 100 null model networks (shown in the boxplots) generated using the degree-preserving edge swap procedure. This is performed by selecting a pair of edges ($A{\rightarrow}B$) ($C{\rightarrow}D$) and swapping them to give ($A{\rightarrow}D$)($C{\rightarrow}B$). If the resulting edges already exist in the network, another pair of edges is selected instead. Each edge was swapped 10 times to ensure full randomisation. To compute the multilink motif *z*-scores, the null model was constructed by randomizing each layer independently.

To identify neurons with high-participation in all of the network layers, the normalized degree-rank product was used. This is computed by ranking neurons in each network layer by their degree in descending order, and scaling to the range [0, 1]. The product is then taken of the ranked degrees in each layer. Thus, if a neuron had the highest degree in each of the network layers, it would have a degree product of 1.

*Clustering coefficient*

The measure of clustering described here is the global clustering, also known as *transitivity,* given in (Newman, 2003; Rubinov and Sporns, 2010; Wasserman and Faust, 1994), which measures the ratio of triangles to triples (where a triple is a single node with edges running to an unordered pair of others, and a triangle is a fully-connected triple). For a directed network, this is equivalent to:

$$T = \frac{\sum_{i \in N} t_i}{\sum_{i \in N}[(k_i^{out} + k_i^{in})(k_i^{out} + k_i^{in} - 1) - 2\sum_{j \in N} A_{ij}A_{ji}]}$$

where $A$ is the adjacency matrix, $N$ is the number of nodes, $k^{out}$ and $k^{in}$ are the out-degree and in-degree, and $t_i$ is the number of triangles around a node:

$$t_i = \frac{1}{2} \sum_{j,h \in N} (A_{ij} + A_{ji})(A_{ih} + A_{hi})(A_{jh} + A_{hj})$$

*Characteristic path length*

To obtain the characteristic path length of a network, the geodesic (i.e. minimum) distance, $d$, between each pair of nodes $i, j,$ is first computed:

$$d_{ij} = \sum_{A_{uv} \in g(i,j)} A_{uv}$$

where $g(i,j)$ returns the geodesic path between nodes *i* and *j*. The characteristic path length is then given:

$$L = \frac{1}{n} \sum_{i \in N} \frac{\sum_{j \in N, i \neq j} d_{ij}}{n-1}$$

*Modularity*

The modularity $Q$ is determined by first subdividing the network into non-overlapping modules $c$ to maximise within-module connectivity and minimise between-module connectivity (Leicht and Newman, 2008). The modularity then gives the proportion of edges that connect to nodes within the same module:

$$Q = \frac{1}{M} \sum_{i,j \in N} \left( A_{ij} - \frac{k_i^{in} k_j^{out}}{M} \right) \delta(c_i, c_j)$$

where $c_i$, $c_j$ are the modules respectively containing nodes $i, j$; $M$ is the number of edges, and $\delta$ is the Kronecker delta function:

$$\delta(x, y) = \begin{cases} 1 & \text{if } x = y \\ 0 & \text{if } x \neq y \end{cases}$$

*Assortativity*

The assortativity of a network is the correlation between the degrees of nodes on either side of a link. This is given by Newman (Newman, 2002) as:

$$R = \frac{M^{-1} \sum_{ij \in E} k_i^{out} k_j^{in} - \left[ M^{-1} \sum_{ij \in E} \frac{1}{2}(k_i^{out} + k_j^{in}) \right]^2}{M^{-1} \sum_{ij \in E} \frac{1}{2}\left([(k_i^{out})^2 + (k_j^{in})^2]\right) - \left[ M^{-1} \sum_{ij \in E} \frac{1}{2}(k_i^{out} + k_j^{in}) \right]^2}$$

*Reducibility*

Structural reducibility measures the uniqueness of layers by comparing the relative Von Neumann entropies. The larger the relative entropy, the more distinguishable the layer. Formally, the Von Neumann entropy for a layer is

given:

$$H = -\sum_{i}^{N} \lambda_i^{[\alpha]} \log_2 \lambda_i^{[\alpha]}$$

where $\lambda_i^{[\alpha]}$ are the eigenvalues of the Laplacian matrix associated to layer $A^{[\alpha]}$. Full details are given in (De Domenico et al., 2015a). The Ward hierarchical clustering method (Ward, 1963) was used to visualise layer similarity.

*Reciprocity*

Reciprocity is the fraction of reciprocal edges in the network:

$$r = \frac{|E^{\leftrightarrow}|}{M}$$

where $M$ is the number of edges, and $|E^{\leftrightarrow}|$ is the number of reciprocal edges:

$$|E^{\leftrightarrow}| = \sum_{i \neq j} A_{ij} A_{ji}$$

*Rich-club coefficient*

The rich club phenomenon is the tendency for high-degree nodes in a network to form highly-interconnected communities (Colizza et al., 2006; Zhou and Mondragon, 2004). Such communities can be identified by creating subnetworks for each degree level $k$, where nodes with a degree $\leq k$ are removed, and computing the rich club coefficient $\Phi(k)$ for each subnetwork. This is the ratio of remaining connections $M_k$ to the maximum possible number of connections. For a directed network with no self-connections, where $N_k$ is the number of remaining nodes, this is given by:

$$\Phi(k) = \frac{M_k}{N_k(N_k - 1)}$$

Thus, a fully-connected subnetwork at a given degree $k$ has a rich club coefficient $\Phi(k) = 1$. To normalise the rich club coefficient, we computed the average values for 100 random networks $\langle\Phi_{random}(k)\rangle$:

$$\Phi_{norm}(k) = \frac{\Phi(k)}{\langle\Phi_{random}(k)\rangle}$$

We used the same threshold previously used in determining the wired rich club of *C. elegans* (Towlson et al., 2013), defined a rich club to exist where $\Phi_{norm}(k) \geq 1 + 1\sigma$, where $\sigma$ is the Standard Deviation of $\Phi_{random}(k)$.

*Multilink motifs*

Multilink motif analysis considers the full range of possible link combinations that can exist between any two nodes across all layers of a network, and is based on the concept of multilinks as described in (Bianconi, 2013; Boccaletti et al., 2014; Menichetti et al., 2014) Due to the conceptual and structural similarity between monoamine layers (see *reducibility*), we limited our analysis to three layers: synaptic, gap junction, and monoamine (see SI for neuropeptides), giving a total of 20 possible multilink motifs. Instances of each motif were recorded by simultaneously traversing the three network layers. This was also conducted for 100 randomized three-layer networks, generated by rewiring each of the real networks individually using the same randomisation procedure described above. These random networks were used to calculate motif *z*-scores and *p*-values for the actual network.

*Software*

Network measures were computed in MATLAB (v8.5, The MathWorks Inc.,

Natick, MA) using the Brain Connectivity Toolbox (Rubinov and Sporns, 2010) and MATLAB/Octave Networks Toolbox (Bounova and de Weck, 2012). Reducibility analysis, clustering, and multilayer plots were computed in MuxViz (De Domenico et al., 2015b). Reducibility is based on the algorithm described in (De Domenico et al., 2015a), and layer similarity was visualized using the Ward hierarchical clustering method (Ward, 1963). Hive plots were generated using the custom hiveplotter function written in Python (Python Software Foundation. Python Language Reference, v3.5). Additional network visualisations were created using Cytoscape (Shannon et al., 2003) and Dia (https://wiki.gnome.org/Apps/Dia/).

**Acknowledgements**

We thank the MRC and Wellcome Trust for funding, and Oliver Hobert, Niels Ringstad, and the Caenorhabditis Genetics Center for strains.

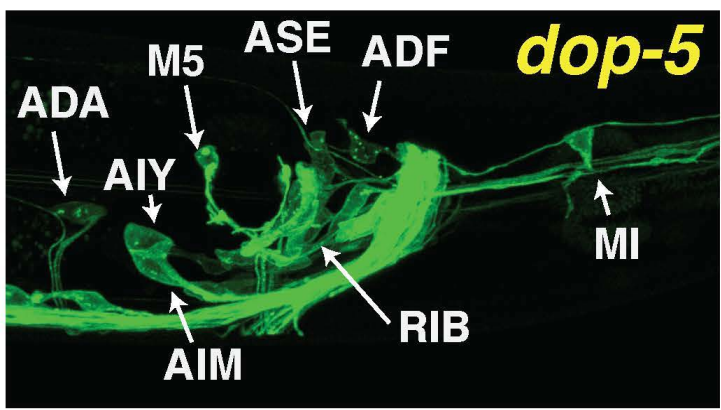
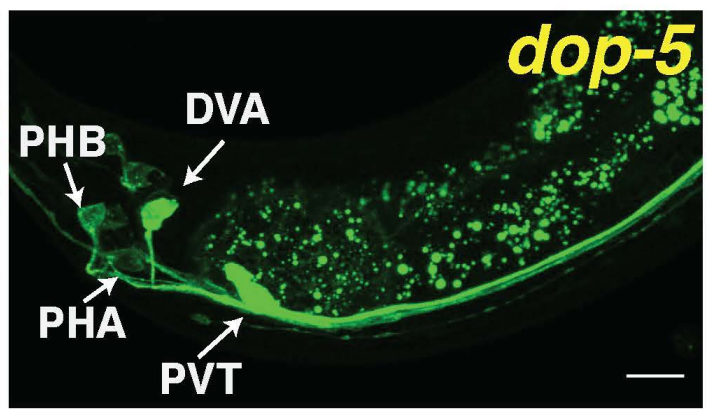
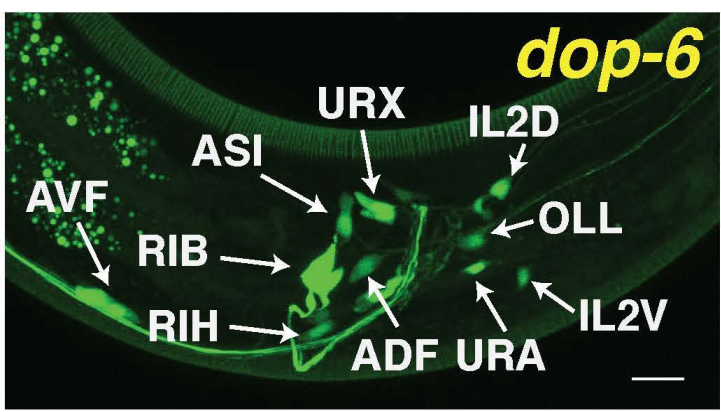
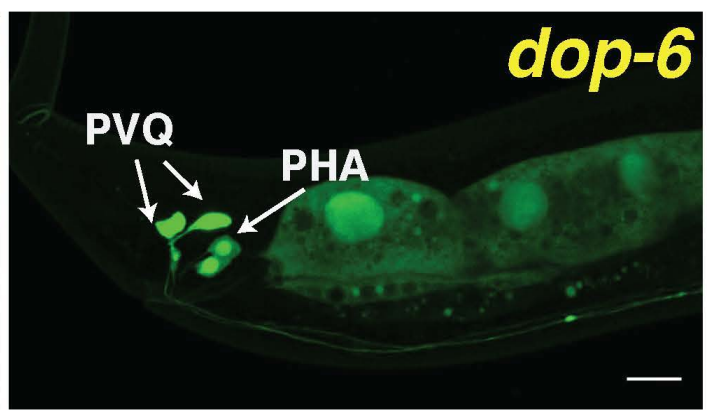
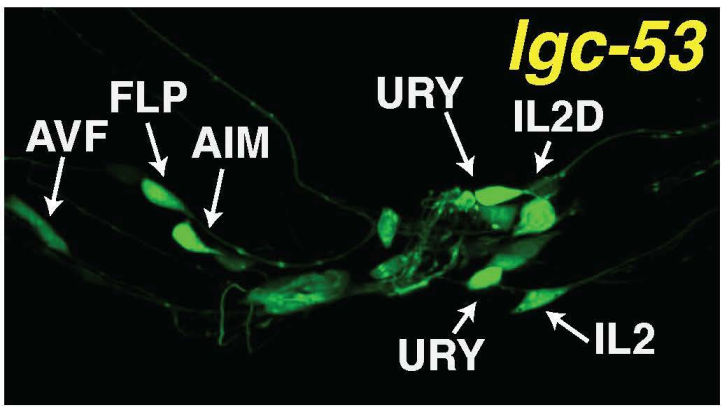
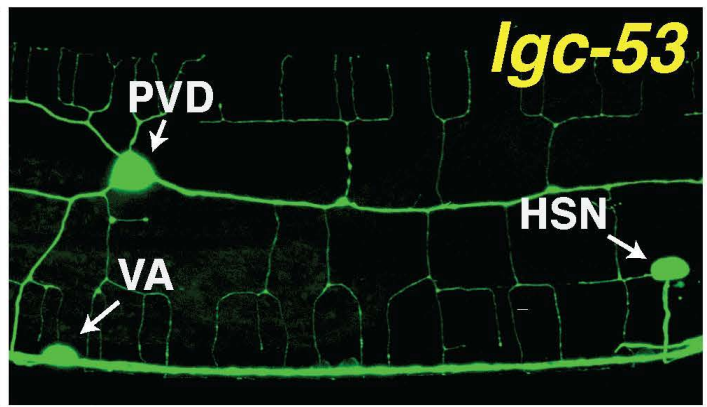

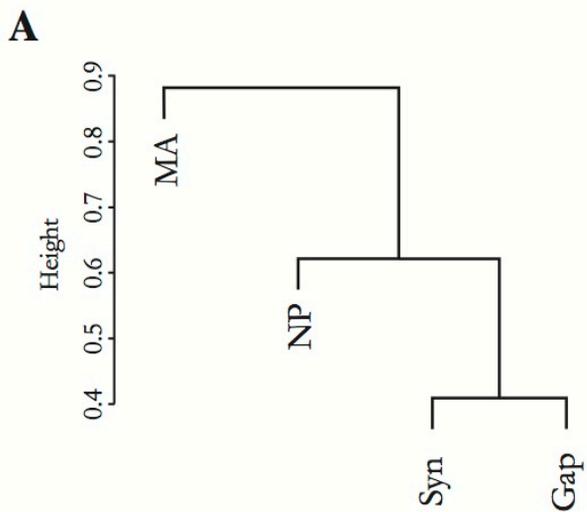

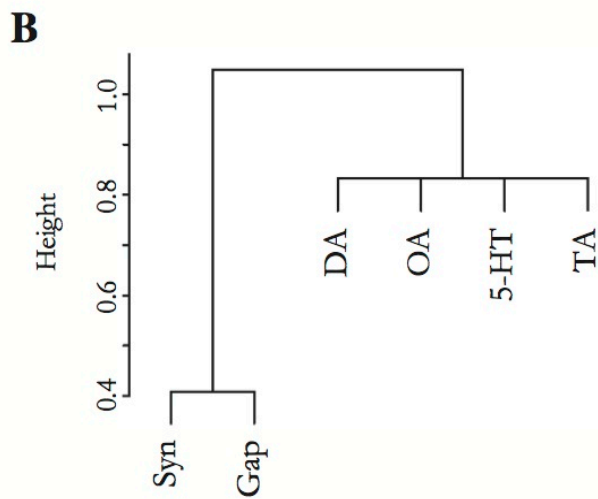

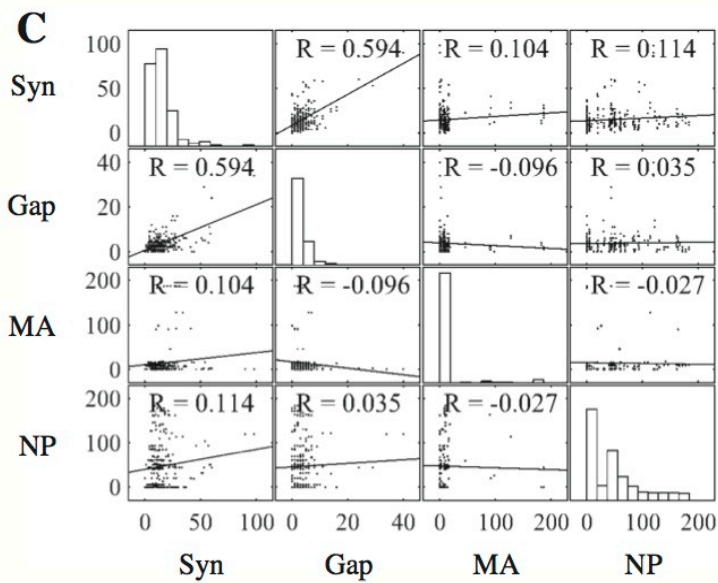

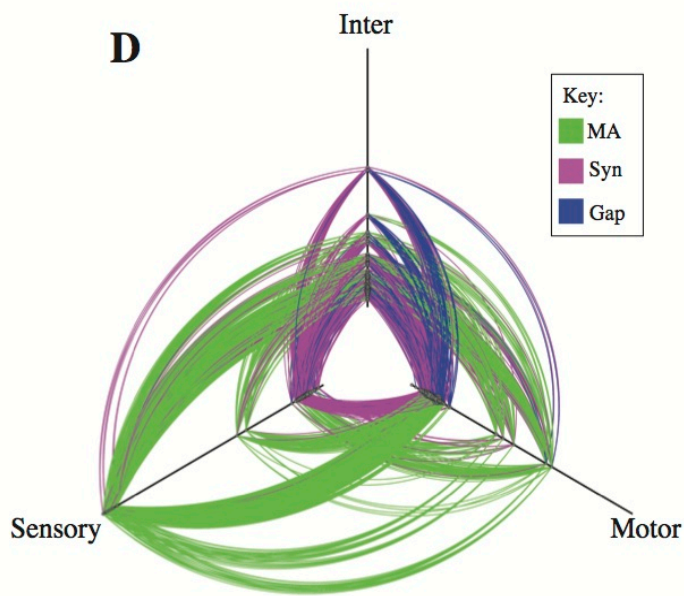

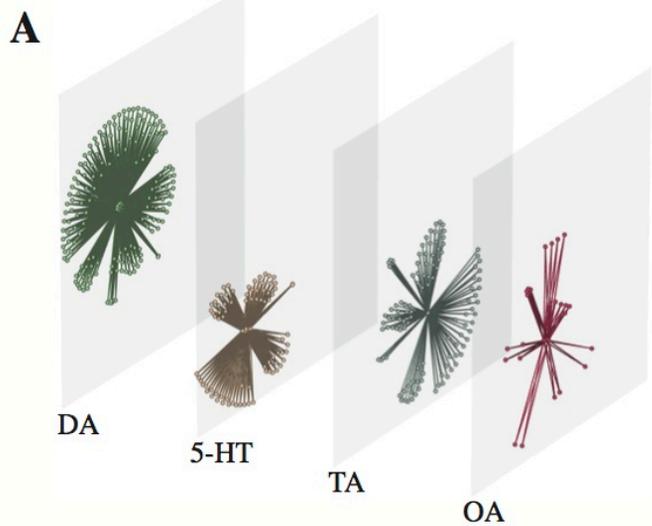
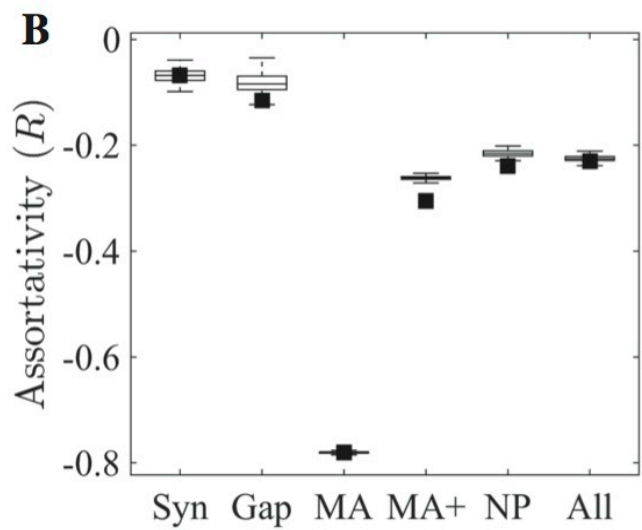
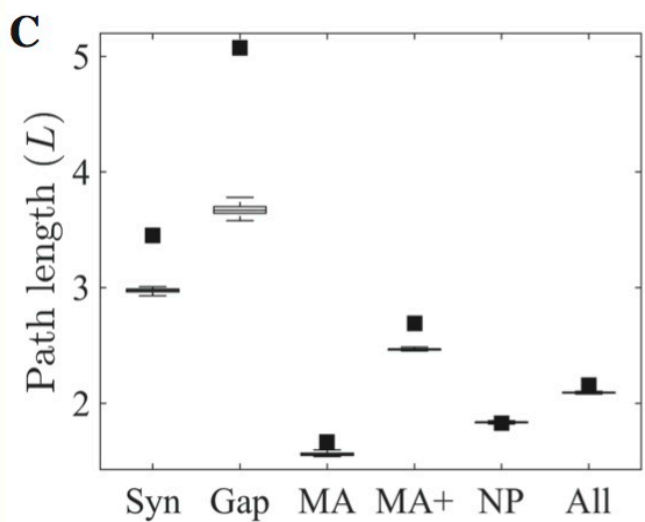
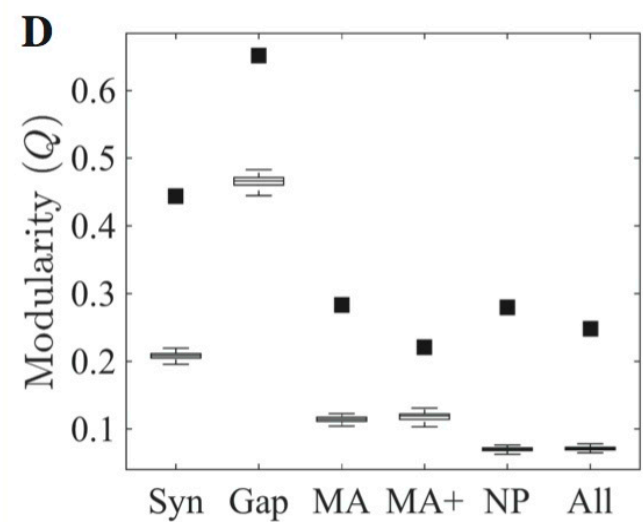
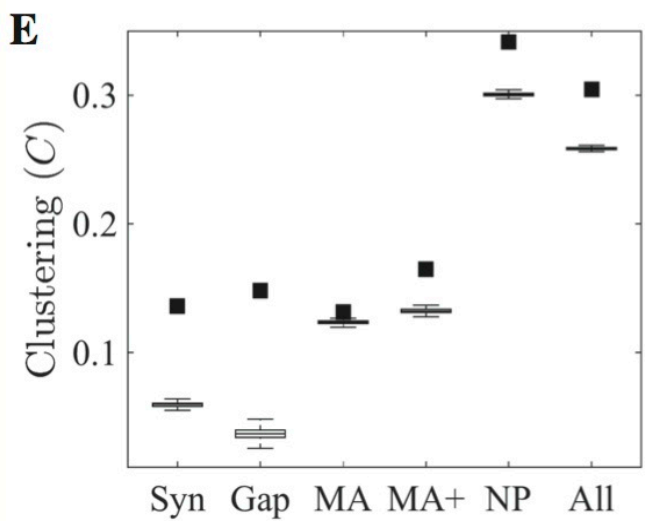
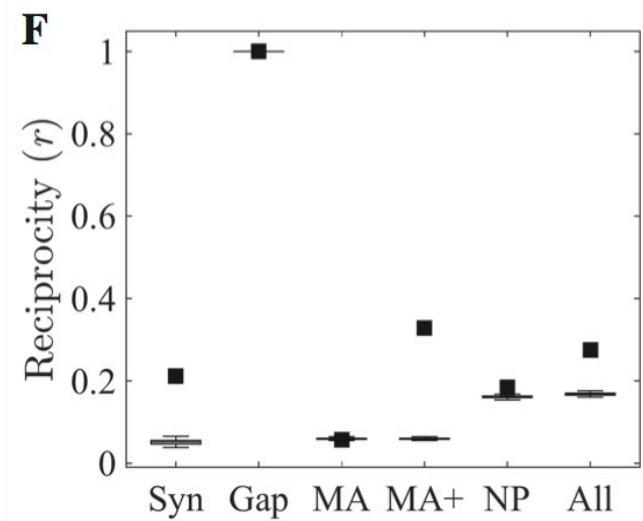

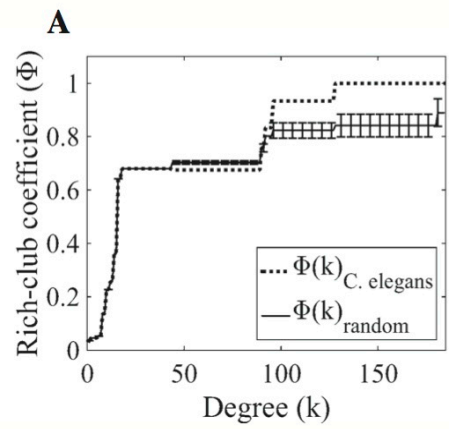 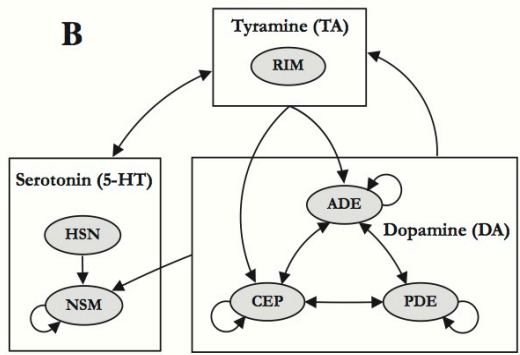 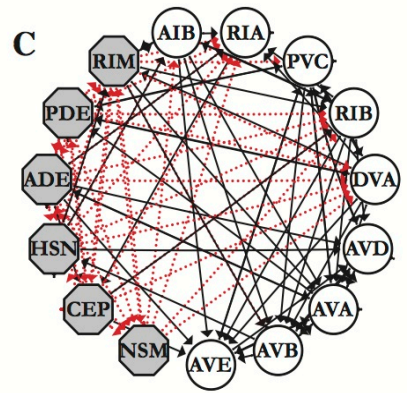

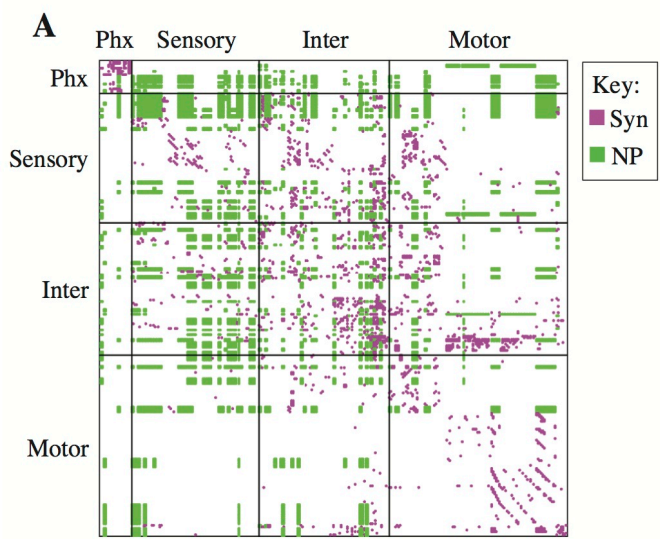
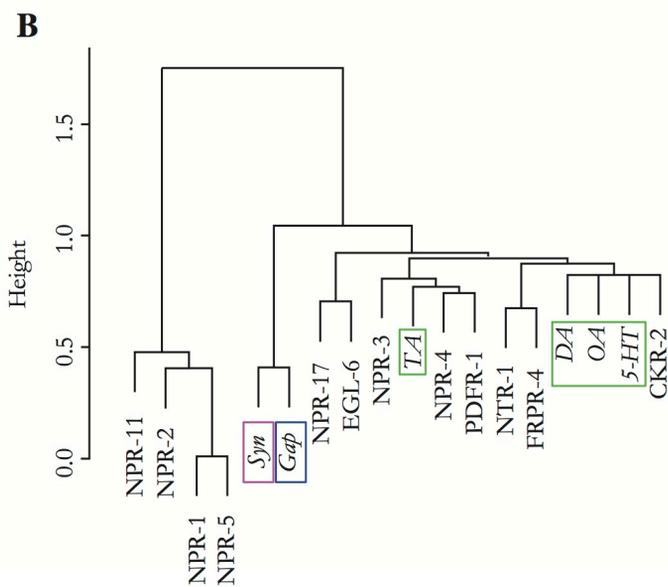
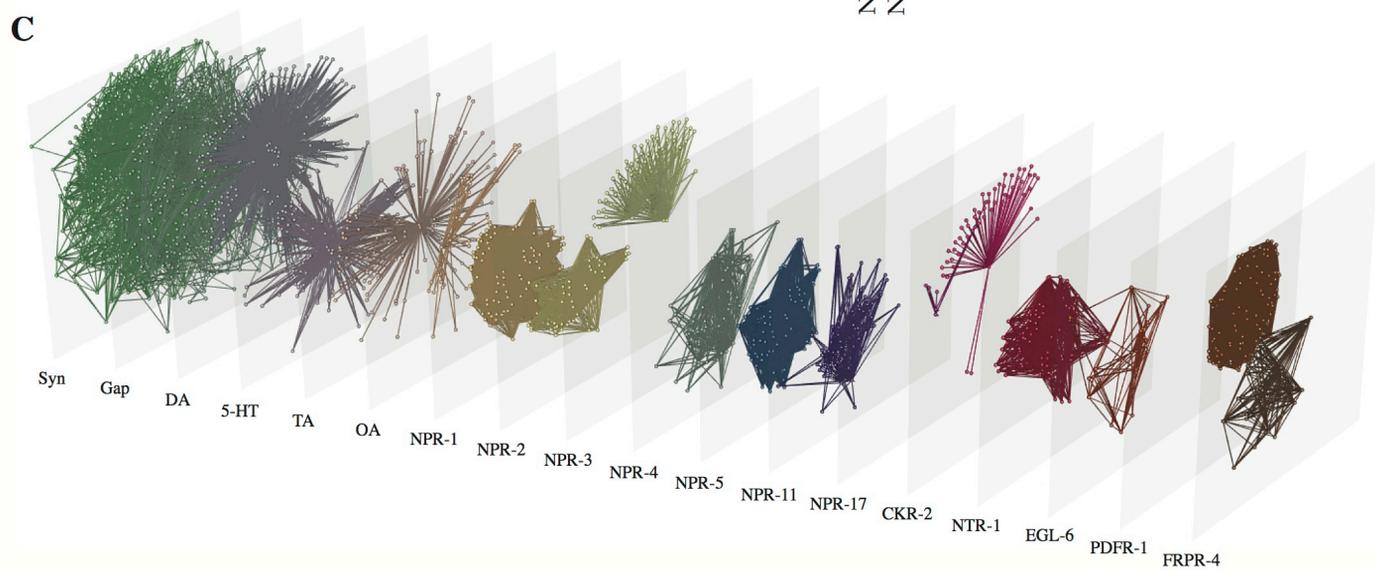

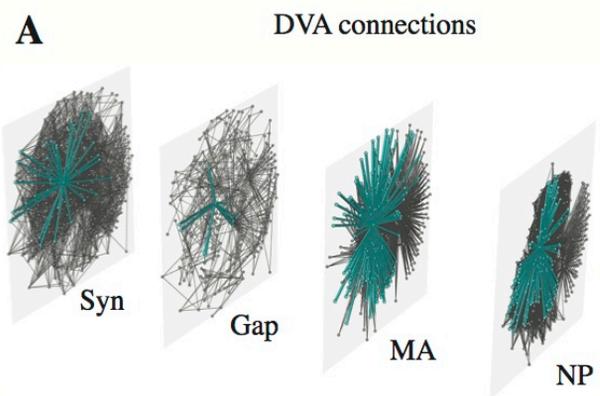
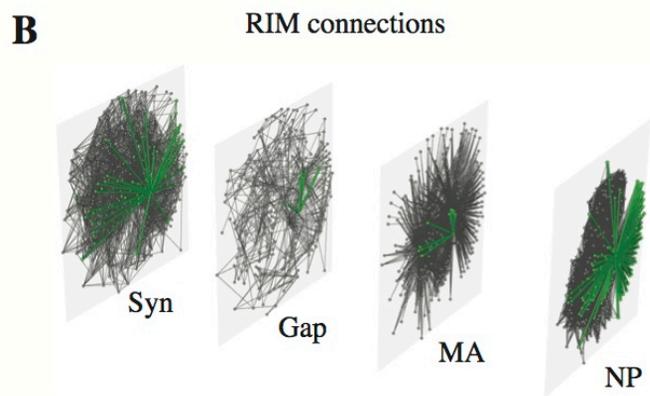
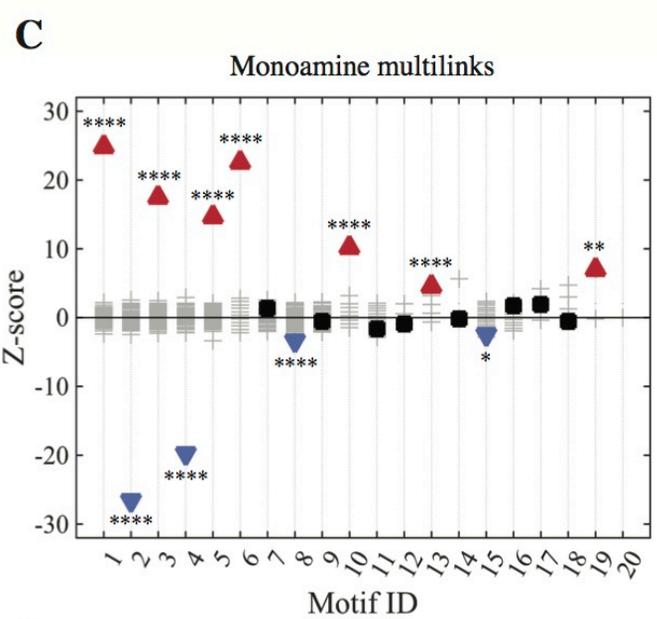
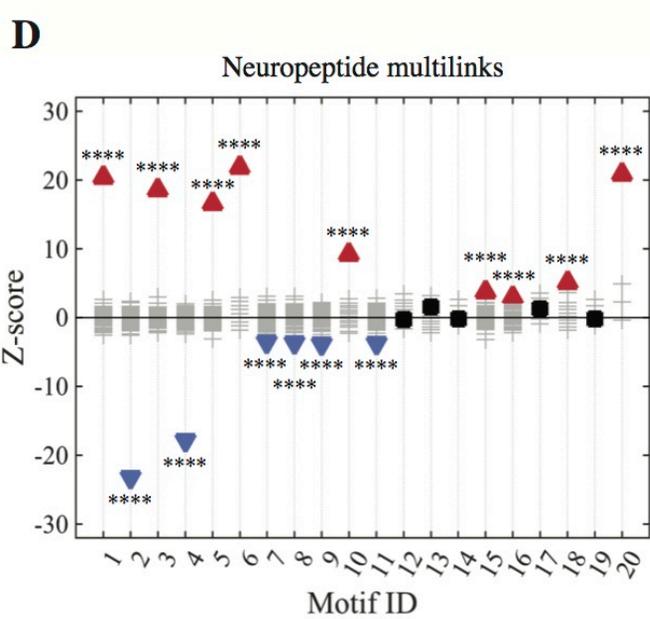
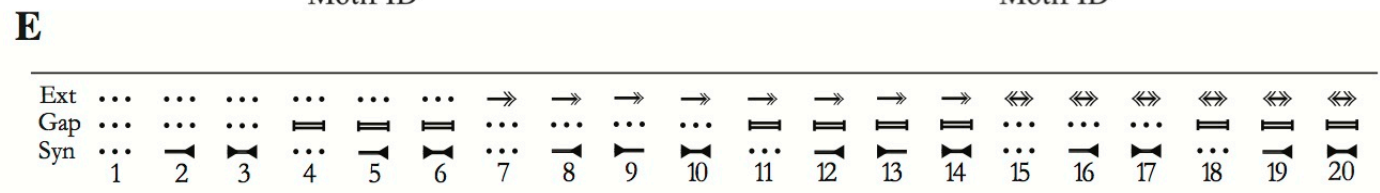

# Supplemental Information

## Supplemental Tables

**Table S1:** Serotonin (5-HT) expressing cells. Cells with weak or conditional expression are marked †

| Marker | WormBase ID | Neurons | Reference |
|---|---|---|---|
| *tph-1* | Expr959 | RIH†, AIM†, ADF, NSM, HSN | (Sze et al., 2000) |
|  | Expr12176 | ASG† | (Pocock and Hobert, 2010) |
| *mod-5* | Expr9350 | AIM†, NSM, ADF, RIH† | (Jafari et al., 2011) |

**Table S2:** Dopamine (DA) expressing cells

| Marker | WormBase ID | Neurons | Reference |
|---|---|---|---|
| *cat-2* | Expr2619 | ADE, PDE, CEP | (Suo et al., 2003) |
| *dat-1* | Expr8327 | ADE, PDE, CEP | (McDonald et al., 2007) |

**Table S3:** Octopamine (OA) & tyramine (TA) expressing cells. ★RIC is excluded from the TA network due to co-expression of *tbp-1* which converts TA to OA

| Marker | WormBase ID | Neurons | Reference |
|---|---|---|---|
| **Octopamine** *tbh-1* | Expr3721 | RIC | (Alkema et al., 2005) |
| **Tyramine** *tdc-1* | Expr3722 | RIM, RIC★ | (Alkema et al., 2005) |

**Table S4:** Serotonin (5-HT) receptor expression patterns

| Marker | WormBase ID | Neurons | Reference |
|---|---|---|---|
| *ser-1* | Expr7825 | RIA, RIC, PVT, DVC, URY | (Dernovici et al., 2007) |
|  | Expr8282 | PVQ | (Carnell et al., 2005) |
|  | Expr3962 | RMD, RMF, RMH | (Xiao et al., 2006) |

| Marker | WormBase ID | Neurons | Reference |
|---|---|---|---|
| *ser-4* | Expr2710 | RIB, PVT, DVC, DVA, RIS | (Tsalik et al., 2003) |
|  | Expr10554 | AIB, NSM | (Gurel et al., 2012) |
|  | N/A | M1, RIM | (Shyn, 2003) |
| *ser-5* | Expr12174 | ASH, AWB | (Hapiak et al., 2009) |
|  | Expr12172 | AVJ | (Cunningham et al., 2012) |
| *ser-7* | Expr3759 | MC, M2, M3, M4, M5, I2, I3, I4, I6 | (Hobson et al., 2006) |
| *mod-1* | Expr10023 | RIM, RID, RIC, AIZ, AIY, AIB, AIA | (Li et al., 2012) |
|  | Expr10553 | RM, DD, VD | (Gurel et al., 2012) |

**Table S5:** Octopamine (OA) receptor expression patterns

| Marker | WormBase ID | Neurons | Reference |
|---|---|---|---|
| *octr-1* | Expr7846 | ASH, ASI, AIY, ADE, CEP | (Wragg et al., 2007) |
| *ser-3* | Expr8275 | PVQ, PHB, PHA, SIA | (Suo et al., 2006) |
|  | Expr10640 | ASH | (Mills et al., 2012) |
| *ser-6* | Expr10641 | AWB, ASI, ADL | (Mills et al., 2012) |
|  | Expr11709 | RIC, SIA | (Yoshida et al., 2014) |

**Table S6:** Dopamine (DA) receptor expression patterns

| Marker | WormBase ID | Neurons | Reference |
|---|---|---|---|
| *dop-1* | Expr2882 | AUA, RIM, ALM, RIB, PLM, PHC | (Sanyal et al., 2004) |
|  | Expr2708 | AVM, ALN, PVQ, PLN, RIS | (Tsalik et al., 2003) |
|  | Expr3047 | PVD, VA, VB, AS, DA, DB | (Chase et al., 2004) |
| *dop-2* | Expr2618 | ADE, PDE, CEP | (Suo et al., 2003) |
|  | Expr2709 | RID, RIA, PDA, SIB, SIA | (Tsalik et al., 2003) |
| *dop-3* | Expr3048 | PVD, VA, VB, AS, DA, DB, DD, VD | (Chase et al., 2004) |
|  | Expr7939 | ASE | (Etchberger et al., 2009) |

|  | Expr8667 | RIC, SIA | (Suo et al., 2009) |
|  | Expr11452 | NSM | (Zhang et al., 2014) |
|  | Expr12177 | ASK | (Ezak and Ferkey, 2010) |
| *dop-4* | Expr3687 | AVL, ASG, PQR, I2, I1, CAN | (Sugiura et al., 2005) |
| *dop-5* | Expr7939 | ASE | (Etchberger et al., 2009) |
|  | N/A | MI, M5, BDU, RIB, PHA, PHB, DVA, AIM, ADA, AIY, PVT | N/A |
| *dop-6* | Expr11993 | OLL | (Smith et al., 2010) |
|  | N/A | RIB, ASI, PHA, IL2, PVQ, URA, AVF, ADF, RIH, URX | N/A |
| *lgc-53* | N/A | HSN, PVD, CAN, IL2, PVPR, VA, AIM, FLP, AVF, URY | N/A |

**Table S7:** Tyramine (TA) receptor expression patterns

| Marker | WormBase ID | Neurons | Reference |
| --- | --- | --- | --- |
| *ser-2* | Expr2707 | BDU, AVH, AUA, ALN, RID, RIC, AIZ, RIA, AIY, PVT, PVD, PVC, OLL, NSM, LUA, DVA, DA09, CAN, SIA, SDQ, SAB, RME | (Tsalik et al., 2003) |
|  | Expr3206 | PVD | (Rex et al., 2004) |
|  | Expr10758 | VD | (Donnelly et al., 2013) |
| *tyra-2* | Expr3415 | ASI, ASH, ASG, ASE, ALM, PVD, NSM, MC, CAN | (Rex et al., 2005) |
| *tyra-3* | Expr11003 | BAG, AWC, AUA, ASK, AIM, AFD, ADL, OLQ, CEP, SDQ | (Bendesky et al., 2011) |
|  | Expr6415 | PVT | (Hunt-Newbury et al., 2007) |
|  | Expr12173 | ADE | (Wragg et al., 2007) |
| *lgc-55* | Expr8613 | AVB, ALN, IL1, HSN, SMD, SDQ, RMD | (Pirri et al., 2009) |
|  | Expr8997 | AVM, ALM | (Ringstad et al., 2009) |

**Table S8:** Neuropeptide expression patterns

| Receptor | WBID | Neurons | Reference |
| --- | --- | --- | --- |
| *flp-1* | Expr3003 | AVK, AVE, AVA, RIG, AIY, AIA, M5, RMG | (Kim and Li, 2004) |
| *flp-4* | Expr3006 | AWC, AVM, ASEL, ADL, PVD, PHB, PHA, NSM, I5, I6, FLP | (Kim and Li, 2004) |
| *flp-5* | Expr3007 | ASE, PVT, M4, I4, I2 RMG | (Kim and Li, 2004) |
| *flp-10* | Expr3011 | AIM, ASI, AUA, BAG, BDU, DVB, PQR, PVR, URX | (Kim and Li, 2004) |
| *flp-13* | Expr3014 | ASE, ASG, ASK, BAG, DD, I5, M3, M5 | (Kim and Li, 2004) |
|  | Expr12005 | ALA | (Nelson et al., 2015) |
| *flp-15* | Expr3015 | PHA, I2 | (Nelson et al., 2015) |
| *flp-17* | Expr3016 | BAG, M5 | (Kim and Li, 2004) |
| *flp-18* | Expr3017 | AVA, AIY, RIG, RIM, M2, M3 | (Kim and Li, 2004) |
| *flp-21* | Expr3020 | ASI, ASH, ASE, ADL, MC, M4, FLP, URA | (Kim and Li, 2004) |
|  | Expr12181 | RMG, ASJ, URX, M2, ASK, ASG, ADF | (Macosko et al., 2009) |
| *nlp-1* | Expr1686 | ASI, AWC, PHB, BDU | (Nathoo et al., 2001) |
|  | Marker88 | HSN | (Karakuzu et al., 2009) |
| *nlp-12* | Expr8057 | DVA | (Janssen et al., 2008a) |
| *ntc-1* | Expr11371 | AVK, RIC, AIZ, AFD, NSM, M5, DVA, DD, VD, VC | (Garrison et al., 2012) |
|  | Expr11368 | ASG | (Beets et al., 2012) |
| *pdf-1* | Expr11002 | AVB, ASK, AIM, AFD, PVT, PVP, PVN, LUA, SIA, SAA, RMG | (Barrios et al., 2012) |
|  | Expr9958 | ASI, RID, ADA, ADE, PQR, PHB, PHA, RME | (Janssen et al., 2008a) |
| *pdf-2 / nlp-3* | Expr9959 | BDU, AVG, AVD, RIM, AQR, RID, AIM, PVT, PVP, PQR, PHB, PHA, RIS | (Janssen et al., 2008a) |
| *nlp-24* | Expr1717 | ASI | (Nathoo et al., 2001) |

**Table S9:** Neuropeptide receptor expression patterns

| Receptor | WBID | Neurons | Reference |
|---|---|---|---|
| *npr-1* | Expr2257 | AUA, ASH, ASG, ASE, AQR, RIG, PQR, PHB, PHA, OLQ, IL2L, IL2R, URX, SMBDL, SMBDR, RMG, RIV, DD, VD, M3, SAADL, SAADR, SDQ | (Coates and de Bono, 2002) |
| *npr-2* | Expr12242 | ADF, AIZ, ASH, FLP, OLQ, PVD, PVQ, SAB | (Luo et al., 2015) |
| *npr-3* | Expr2766 | AS, DA, DB, VA, VB | (Keating et al., 2003) |
| *npr-4* | Expr8975 | BDU, BAG, AVA, PQR, RIV | (Cohen et al., 2009) |
| *npr-5* | Expr8976 | AWB, AWA, AUA, ASK, ASJ, ASI, ASG, ASE, AIA, ADF, PHB, PHA, IL2 | (Cohen et al., 2009) |
| *npr-11* | Expr12179 | AIA, AIY | (Chalasani et al., 2010) |
| *frpr-4* | N/A | RIA, PVM, AVE, I1, DVA | (Nelson et al., 2015) |
| *npr-17* | Expr12182 | AVG, ASI, PVP, PVQ, PQR | (Harris et al., 2010) |
| *ckr-2* | Expr10065 | AIY | (Wenick and Hobert, 2004) |
|  | Expr12178 | AS, DA, DB, VA, VB | (Hu et al., 2011) |
| *ntr-1* | Expr11372 | ASH, RIC, ADL, ADF, PVW, PVR, PVQ, I2 | (Garrison et al., 2012) |
|  | Expr11369 | BDU, ASE, PQR | (Beets et al., 2012) |
| *egl-6* | Expr8338 | HSN, DVA, SDQ | (Ringstad and Horvitz, 2008) |
| *pdfr-1* | Expr10592 | AVM, AVD, RIF, ALM, PVW, PVQ, PVM, PVC, PQR, PLM, PHA, OLL, DB2, URY, URX, RME, AVF | (Barrios et al., 2012) |
|  | Expr8177 | PHB, OLQ, I1 FLP | (Janssen et al., 2008a) |

**Table S10:** Neuropeptide receptor-ligand binding. ★No EC50 value reported for NPR- 11/NLP-1; strong biological activity seen in the micromolar range

| Receptor | Ligand | $EC_{50}$ | Reference |
|---|---|---|---|
| **NPY(npr) / RFamide receptor group** | | | |
| NPR-1 | FLP-18 | ~ 100 nM | (Rogers et al., 2003) |
| | FLP-21 | 2.5 nM | (Kubiak et al., 2003a) |
| NPR-2 | FLP-21 | 34.4 nM | (Ezcurra et al., 2016) |
| NPR-3 | FLP-15 | 162-599 nM | (Kubiak et al., 2003b) |
| NPR-4 | FLP-1 | 0.4-9 μM | (Geary et al., 2002) |
| | FLP-4 | 5-80 nM | (Geary et al., 2002) |
| | FLP-18 | 5 nM-1.2 μM | (Cohen et al., 2009) |
| NPR-5 | FLP-18 | 13.3-117.2 nM | (Kubiak et al., 2008) |
| | FLP-21 | 267 nM | (Kubiak et al., 2008) |
| NPR-11 | FLP-1 | 1-8 μM | (Geary et al., 2002) |
| | FLP-5 | 1-8 μM | (Geary et al., 2002) |
| | FLP-18 | 180-800 nM | (Geary et al., 2002) |
| | FLP-21 | 1-10 nM | (Geary et al., 2002) |
| | NLP-1 | 1-100 μM? | (Chalasani et al., 2010) |
| FRPR-4 | FLP-13 | 67-541 nM | (Nelson et al., 2015) |
| **Somatostatin / Urotensin II receptor group** | | | |
| NPR-17 | FLP-24 | 0.1-1 μM | (Cheong et al., 2015) |
| **Gastrin / CCK-like receptor group** | | | |
| CKR-2 | NLP-12 | 15-30 nM | (Janssen et al., 2008b) |
| **Vasopressin-like receptor group** | | | |
| NTR-1 | NTC-1 | 19 nM | (Beets et al., 2012) |
| **Neurotensin / TPH-like receptor group** | | | |
| EGL-6 | FLP-10 | 11 nM | (Ringstad and Horvitz, 2008) |
| | FLP-17 | 1-28nM | (Ringstad and Horvitz, 2008) |
| **Class B / Secretin receptor group** | | | |
| PDFR-1 | PDF-1 | 0.4-5 μM | (Janssen et al., 2008a) |
| | PDF-2 / NLP-37 | 114 nM | (Janssen et al., 2008a) |

**Supplemental References**